# Neural Networks Search for Charged Higgs Boson of Two Doublet Higgs Model at the Hadrons Colliders

Nady Bakhet[1,2], Maxim Yu Khlopov[3,4], Tarek Hussein[1,2]

[1]Department of Physics, Cairo University, Giza, Egypt
[2]Egyptian Network of High Energy Physics - ASRT, Cairo, Egypt
[3]APC Laboratory, IN2P3/CNRS, Paris, France
[4]National Research Nuclear University "MEPHI" (Moscow Engineering Physics Institute), Moscow, Russia

**ABSTRACT**

In this work, we present an analysis of a search for charged Higgs boson in the context of Two Doublet Higgs Model (2HDM) which is an extension of the Standard Model of particles physics. The 2HDM predicts by existence scalar sector with new five Higgs bosons, two of them are electrically charged and the other three Higgs bosons are neutral charged. Our analysis based on the Monte Carlo data produced from the simulation of 2HDM with proton antiproton collisions at the Tevatron $\sqrt{s} = 1.96\ TeV$ (Fermi Lab) and proton proton collisions at the LHC $\sqrt{s} = 14\ TeV$ (CERN) with final state includes electron, muon, multiple jets and missing transverse energy via the production and decay of the new Higgs in the hard process $pp(\bar{p}) \to t\bar{t} \to H^+ b H^- \bar{b} \to \tau^+ \nu_\tau b \tau^- \bar{\nu}_\tau \bar{b} \to w^+ \bar{\nu}_\tau \nu_\tau b w^- \nu_\tau \bar{\nu}_\tau \bar{b} \to \mu^+ \nu_\mu \bar{\nu}_\tau \nu_\tau b e^- \bar{\nu}_e \nu_\tau \bar{\nu}_\tau \bar{b}$ where the dominant background (electrons and muons) for this process comes from the Standard Model processes via the production and decay of top quark pair. We assumed that the branching ratio of charged Higgs boson to tau lepton and neutrino is 100%. We used the Artificial Neural Networks (ANNs) which is an efficient technique to discriminate the signal of charged Higgs boson from the SM background for charged Higgs boson masses between 80 GeV and 160 GeV. Also we calculated the production cross section at different energies, decay width, branching ration and different kinematics distribution for charged Higgs boson and for the final state particles.

**Keywords**: 2HDM, Neural Networks, Pythia8, MadGraph5, CalcHep, ROOT

# 1 Introduction

After the new discovery of the Standard Model Higgs boson at CERN's Large Hadron Collider LHC on 2012 [1,2], it is now time to test possible many extensions of the Standard Model (SM) using Monte Carlo simulation techniques and different computational tools of HEP. The Standard Model (SM) does not contain any elementary charged scalar particle; the observation of a charged Higgs boson would indicate new physics beyond the Standard Model. In the Standard Model of the electroweak interactions [3] the masses of both bosons and fermions are explained by the Higgs mechanism [4]. This implies the existence of new one doublet of complex scalar fields which, in turn, leads to a single neutral scalar Higgs boson. One of the simplest ways to extend the scalar sector of the Standard Model is to add one more complex doublet to the model. Some extensions to the Standard Model contain more than one Higgs doublet [5] and predict Higgs bosons which can be lighter than the Standard Model Higgs. The models with two complex Higgs doublets predict two charged Higgs bosons $H^{\pm}$ which can be pair-produced in proton proton collisions (LHC) and proton antiproton collisions (Tevatron) such these models as Two-Doublet Higgs Model (2HDM) [6] and Minimal SuperSymmetric Model (MSSM). The two Higgs doublet model (2HDM) can provide additional CP-violation coming from the scalar sector and can easily originate dark matter candidates, also the Minimal SuperSymmetric Model (MSSM) predicts two doublet Higgs. The 2HDMs have a richer particle spectrum with two charged and three neutral Higgs Bosons. All neutral Higgs Boson could in principle be the scalar discovered at the LHC [7-9]. The SM picks up the ideas of local gauge invariant and Spontaneous Symmetry Breaking (SSB) to implement a Higgs mechanism. The symmetry breaking is implemented by introducing a scalar doublet

$$\Phi = \begin{pmatrix} \phi^+ \\ \phi^0 \end{pmatrix} = \begin{pmatrix} \phi_1 + i\phi_2 \\ \phi_3 + i\phi_4 \end{pmatrix}$$

In order to induce the SSB the doublet should acquire a VEV different from zero

$$<\Phi> = \begin{pmatrix} 0 \\ v/\sqrt{2} \end{pmatrix}$$

The 2HDM introduced a new Higgs doublet so the Higgs sector includes two Higgs doublets with the same quantum numbers.

$$\Phi_1 = \begin{pmatrix} \phi_1^+ \\ \phi_1^0 \end{pmatrix} \text{ And } \Phi_2 = \begin{pmatrix} \phi_2^+ \\ \phi_2^0 \end{pmatrix}$$

With hypercharges $Y_1 = Y_2 = 1$ both doublets could acquire VEV

$$<\Phi_1> = \frac{v_1}{\sqrt{2}} \quad \text{and} \quad <\Phi_2> = \frac{v_2}{\sqrt{2}} e^{i\theta}$$

In the next section we will present an analysis for signatures of the charged Higgs boson in the mass range 80–160 GeV using top quark pair events with a leptonically decaying in the context of 2HDM using Monte Carlo simulation programs and Artificial Neural Network (ANNs) at the LHC $\sqrt{s} = 14\,TeV$ with proton-proton collisions (CMS and ATLAS detectors) and the Tevatron $\sqrt{s} = 1.96\,TeV$ with proton-antiproton collisions (CDF and D0 detectors) with electron, muons, multiple jets and missing transverse energy in the final state, we assumed that the branching ratio of the charged Higgs boson to a τ lepton and a neutrino is 100%.

## 2 The Analysis

In the 2HDM, the scalar sector has two charged Higgs bosons and three neutral Higgs bosons. In current section we will present the results of Monte Carlo Simulation for production and decay the charged Higgs boson at both the LHC $\sqrt{s} = 14$ TeV and at the Tevatron $\sqrt{s} = 1.96$ TeV.

Our search for charged Higgs bosons is based on the following $t\bar{t}$ final states: the dilepton ($\ell\ell$) channel where both charged bosons (W+ or H+) decay into a light charged lepton ($\ell = e\ or\ \mu$) either directly or through the leptonic decay of a $\tau$, the $\tau$+lepton ($\tau\ell$) channel where one charged boson decays to a light charged lepton and the other one to a $\tau$ -lepton decaying hadronically, and the lepton plus jets ($\ell$+jets) channel where one charged boson decays to a light charged lepton and the other decays into hadrons.

The charged Higgs boson is expected to produce via the process $pp(\bar{p}) \to t\bar{t} \to H^+bH^-\bar{b}$ and decay through the decay channels $H^+ \to \tau^+\tau_\nu$, $H^+ \to c\bar{s}$ and other decay channels as shown in figure 4. The process $pp(\bar{p}) \to t\bar{t} \to H^+bH^-\bar{b}$ gives then rise to the signature: $\mu^+\nu_\mu\bar{\nu}_\tau\nu_\tau be^-\bar{\nu}_e\nu_\tau\bar{\nu}_\tau\bar{b}$ .This signature have to be discriminated from the large background of $pp(\bar{p}) \to t\bar{t} \to w^+bw^-\bar{b}$ . The search for pair-produced charged Higgs bosons is performed using Pythia8 Monte Carlo programs [10]. Figure 2 shows the production cross section of the Higgs boson at the LHC $\sqrt{s} = 14$ TeV and at the Tevatron $\sqrt{s} = 1.96$ TeV. It has the maximum value at 140 GeV. The samples of events for $pp(\bar{p}) \to t\bar{t} \to H^+bH^-\bar{b}$ were generated with the MadGraph5/MadEvent matrix elements generators for $M_{H^\pm}$ between 100 GeV and 160 GeV. About 5000 events for the final state are generated at each Higgs mass.

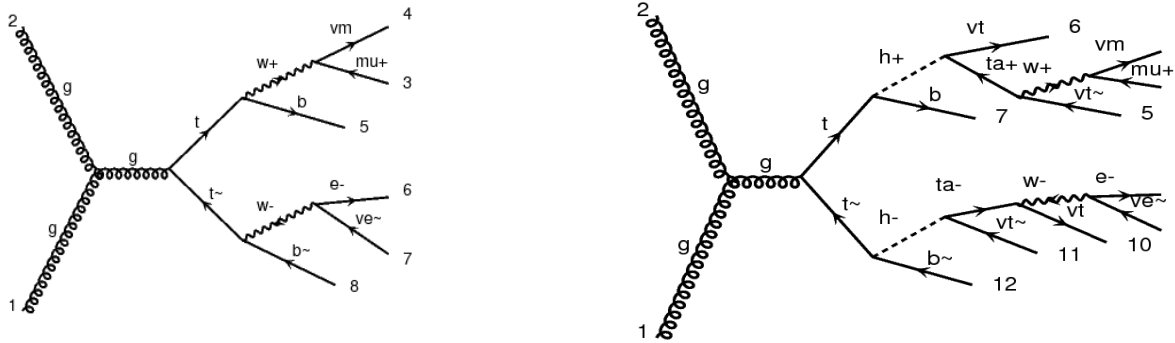

**FIG. 1:** Feynman diagrams for the pair production of $t\bar{t}$ events from gluon fusion where a top quark decays to a charged Higgs boson followed by the decay $H^{\pm} \to \tau\nu$,
(Right) the signal (Left) the background.

For top quark mass 175 GeV may be a source of charged Higgs production. If kinematically allowed, the top quark can decay to $H^+ b$ competing with the Standard Model decay $t \to w^+ b$. This mechanism can provide a larger production rate of charged Higgs and offers high clean signature than that of direct production. The background processes that enter this search include the Standard Model pair production of top quarks $pp(\bar{p}) \to t\bar{t} \to w^+ b w^- \bar{b}$

## 2.1 Production Cross Section of $H^{\pm}$

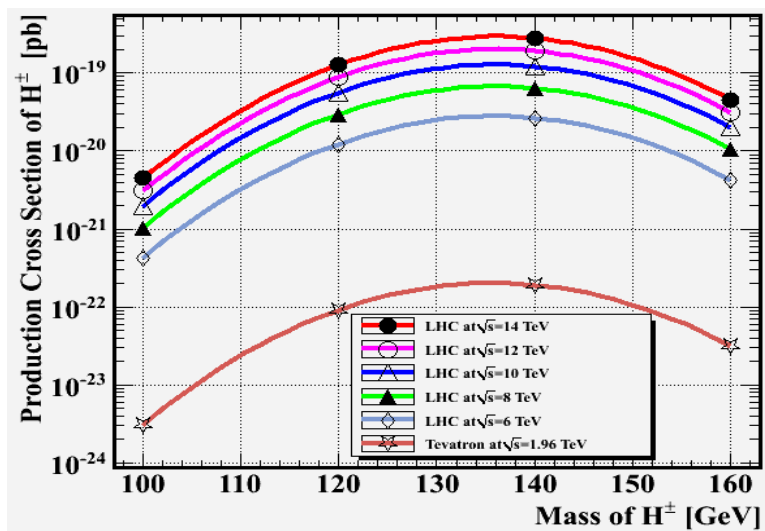

**FIG. 2**: Production Cross section of Charged Higgs boson at the LHC for energies 14,12,10,8 and 6 TeV also at the Tevatron for 1.96 TeV in 2HDM using MadGraph5/Madevent

In the context of the Two Higgs doublet Model, the charged Higgs boson couplings are specified in terms of the electric charge and the weak mixing angle $w_\theta$. The CDF detectors at Fermi Lab has reported measurements of the $t\bar{t}$ production cross section in the $\ell + \not{E}_T + jets + Q$ channels where $\ell$ = e, $\mu$ and where Q = $\ell$ is "dilepton" channel, Q = $\tau$ "lepton+tau", Q = one or more tagged jets ( a jet is determined to be tagged if it shows a displaced secondary vertex, these jets originate from the decay of long lived mesons such as those resulting after the hadronization process of the b quark).

The production cross-section thus depends only on the mass $m_{H^\pm}$. The analyses are not sensitive to the quark flavour, Therefore BR ($H^+ \to \tau^+ \tau_\nu$) = 10% is assumed and $H^+H^-$ pair production leads to the final State $\mu^+ \nu_\mu \bar{\nu}_\tau \nu_\tau b e^- \bar{\nu}_e \nu_\tau \bar{\nu}_\tau \bar{b}$. The signal detection efficiencies and accepted background cross sections are estimated using a variety of Monte Carlo samples. For production and decay of charged Higgs boson at the hadron colliders, we used t → $H^+$b  which is the main production mode at the Large Hadron Collider (LHC) is through top quark decays where the search for a charged Higgs boson is sensitive to the decays of the top quark pairs  pp($\bar{p}$) → $t\bar{t}$ → $H^+ b H^- \bar{b}$ . ATLAS Collaboration upper limits [11] on the production cross section times branching ratio, $\sigma(pp \to tH^\pm + X) \times (H^\pm \to \tau^\pm \nu)$, between 0.76 pb and 4.5 fb, for charged Higgs boson masses ranging from 180 GeV to 1000 GeV and CMS Collaboration upper limits [12] on the production cross section is 0.38–0.026 pb on $\sigma(pp \to \bar{t}(b)H^+) \times B(H^+ \to \tau^+ \nu_\tau)$ for $m_{H^+}$ in the range of 180 to 600 GeV. For the kinematics cuts for electron , muon and Jets. To be the final state of this process is one electron , one muon, 2 jets and missing transverse energy and The tau candidate isolation is based on a cone of $\Delta R = \sqrt{\Delta\varphi^2 + \Delta\eta^2}$=0.5. At the Tevatron, direct production of single charged Higgs is expected to be negligible, and the direct production of $H^+H^-$ via the weak interaction is expected to have a relatively small cross section [13].

The measurements of top quark pair production cross sections $t\bar{t}$ in various channels [14] are sensitive to the decay of top quarks to charged Higgs bosons. The CDF Collaboration reported a search for charged Higgs bosons using different $t\bar{t}$ decay channels with a data set of about $200 pb^{-1}$ [15], resulting in B(t → H$^+$b) <0.4 within the tauonic model. Recently, D0 reported limits on B(t → H$^+$b) for the tauonic and leptophobic models extracted from cross section ratios [16] and for the tauonic model based on a measurement of the $t\bar{t}$ cross section in l+jets channel using topological event information [17].

The Backgrounds W+jets and the signal processes are generated with Madgraph5/Madevent and Pythia8. The eμ events are selected by requiring an electron or muon with $P_T > 10\ GeV$. At least one isolated electron and at least one isolated muon in a cone of radius ΔR = 0.3 around the lepton. The leptons are required to be separated from any selected jet by a distance ΔR = 0.5. The invariant mass of electron-muon pair is required to exceed 15 GeV and the electron and the muon are required to have opposite electric charges. The charged Higgs boson production cross section in the Two Higgs Doublet Model is shown as in figure 2. The search for the fully leptonic final state $H^+H^- \to \tau^+\nu_\tau\tau^-\bar{\nu}_\tau$ is described in [18]. We apply the kinematics cuts on events to identify signal events. The selections events at the LHC $\sqrt{s} = 14$ TeV and the Tevatron $\sqrt{s} = 1.96$ TeV are aimed at charged Higgs boson masses around the expected sensitivity reach of about 120 GeV. The results are used to set upper bounds on the charged Higgs-boson pair production cross section relative to the 2HDM prediction as calculated by MadGraph5/MadEvent. Fully simulated events reconstructed with the Monte Carlo programs and neural networks were used for the background estimates, the design of the selections and the optimization of the selection cuts and the most important background sources come from the decay of $w^+w^-$ bosons. The signal events generated with the Pythia8 Monte Carlo event generator program were simulated for each of the final state for centre of mass energies and for charged Higgs boson masses between 80 and 120 GeV. Leptonic events $w^+w^- \to \mu^+\nu_\mu e^-\bar{\nu}_e$ are rejected by requiring that the momentum of any identified electron or muon be small. The missing energy is required to be greater than 80GeV and the missing mass greater than 70GeV. In order to improve the $w^+w^-$ background rejection, ANNs have been used to construct discriminations. The missing transverse momentum of the event $P_T^{miss}$

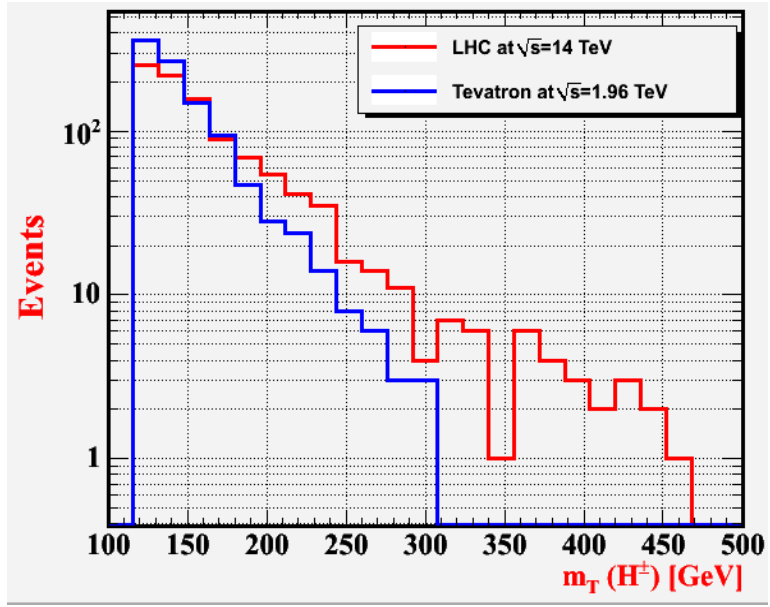

**FIG. 3**: Transverse mass of Production Charged Higgs boson at the LHC at 14 TeV and the Tevatron at 1.96 TeV in 2HDM using MadGraph5, Pythia8

## 2.2 The final state $\mu^+\nu_\mu\bar{\nu}_\tau\nu_\tau b e^-\bar{\nu}_e\nu_\tau\bar{\nu}_\tau\bar{b}$

The final state $\mu^+\nu_\mu\bar{\nu}_\tau\nu_\tau b e^-\bar{\nu}_e\nu_\tau\bar{\nu}_\tau\bar{b}$ is characterized by two jets come from the hadronic decay of the top quark pair with one of the charged Higgs bosons for each. The presence of a light charged Higgs boson would result in a different distribution of $t\bar{t}$ events between different final states than expected in the SM. We select events with e $\mu$ with one isolated high $P_T$ electron and one muon and exactly one or at least two jets. The electron or muon had a small momentum and energy deposition, it was assumed to come from a $\tau$ decay and was therefore tagged as $\tau$ and isolated jets with an energy of at least 5 GeV, at least one and at most five charged particles and no more than ten particles in total were also considered as $\tau$ candidates [19].

## 2.3 Branching Ratios of $H^\pm$

From figure 4, we assume that the charged Higgs boson can decay only to $c\bar{s}$, $\tau\nu_\tau$, $c\bar{b}$, $\mu\nu_\tau$, $\mu\nu_\mu$ and also the charge conjugated decays are implied. The hadronic decay channel $H^\pm \rightarrow cs$ has the highest branching ratio is approximately 50% for all values of charged Higgs masses and the leptonic decay channel $H^\pm \rightarrow \tau\nu_\tau$ is approximately 40% this lead to the possible decay mode for a single top quark $t \rightarrow H^+b$ as shown in figure 5. In current search we used the leptonic decay channel and explicitly set BR($H^\pm \rightarrow \tau\nu_\tau$) = 100% and turn off all other decay channels and evaluated the ratio of BR($t \rightarrow bH^\pm$) as a function of charged Higgs masses. The value of $\Gamma_{H^\pm}$ has little effect on the results as width corrections to the efficiency are small and the relation between the width of the top and the width of the charged Higgs is:

$$\Gamma_t = \frac{\Gamma_{H^\pm}}{1 - \text{BR}(t \rightarrow H^+b)}$$

Within the Standard Model, the top quark decay into a W boson and a b quark occurs with almost 100% probability. The $t\bar{t}$ final state signatures are determined by the W boson decay modes. The decay modes of the charged Higgs boson depend on the ratio of the vacuum expectation values (VEV) of the two Higgs doublets. ATLAS Collaboration limits on the product of branching ratios $B(t \rightarrow bH^\pm) \times (H^\pm \rightarrow \tau^\pm\nu)$ between 0.23% and 1.3% for charged Higgs boson masses in the range 80–160 GeV and CMS Collaboration, is 1.2–0.16% on $B(t \rightarrow bH^+) \times B(H^+ \rightarrow \tau^+\nu_\tau)$ for $m_{H^+}$ in the range of 80 to 160 GeV

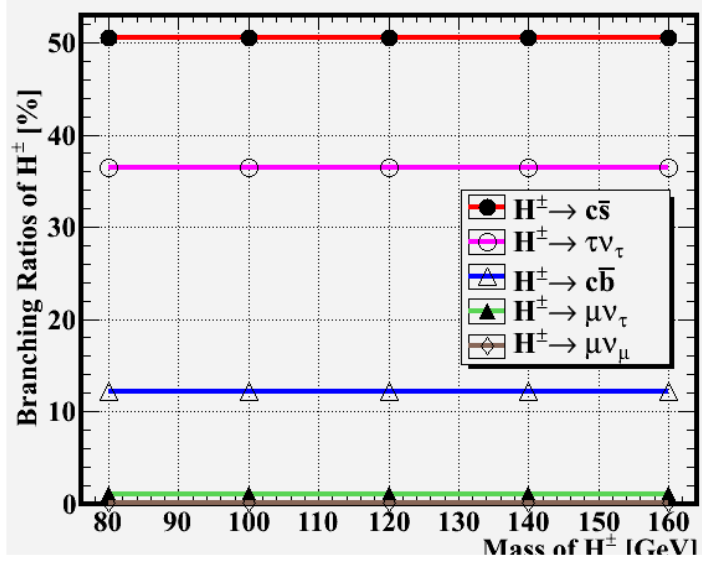

**FIG. 4**: Branching Ratios of all decay channels of Charged Higgs boson as a function of its mass in 2HDM using CalcHep.

In the $\tau^+\nu_\tau b\tau^-\bar{\nu}_\tau\bar{b}$ final state the number of unknowns was higher than the number of constraints and no mass could be estimated.

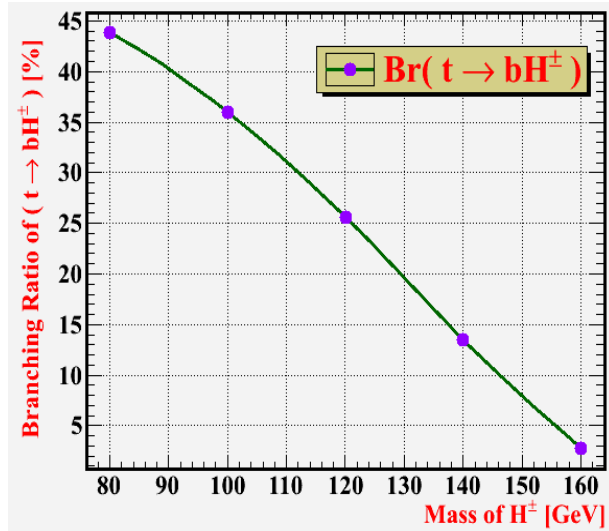

**FIG. 5**: Branching Ratio of decay channel BR( $t \to H^+b$) as a function of mass of Charged Higgs boson in 2HDM using CalcHep.

## 2.4 Decay Width of $H^{\pm}$

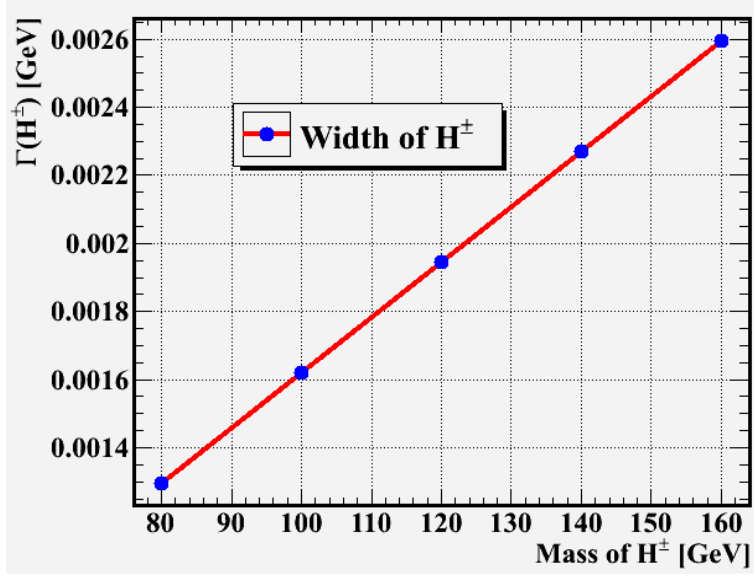

**FIG. 6**: Decay Width of the Charged Higgs boson as a function of its mass.

After final event selection, the signature for $H^+H^- \to \mu^+\nu_\mu\bar{\nu}_\tau\nu_\tau e^-\bar{\nu}_e\nu_\tau\bar{\nu}_\tau$ are electron and muon and large missing energy and momentum and the main backgrounds is the $w^+w^-$ to leptonic decays electron and muon. The $H^+H^-$ signal and the $w^+w^-$ background have similar topologies and the presence of missing neutrinos in the decay of each of the bosons makes the boson mass reconstruction impossible.

The cylindrical coordinates (r, $\phi$) are used in the transverse plane, $\phi$ being the azimuthal angle around the beam pipe. The pseudorapidity $\eta$ is defined in terms of the polar angle $\theta$ as $\eta = -\ln \tan(\frac{\theta}{2})$ where $\theta$ is the angle between the particle three-momentum $\vec{p}$ and the positive direction of the beam axis.

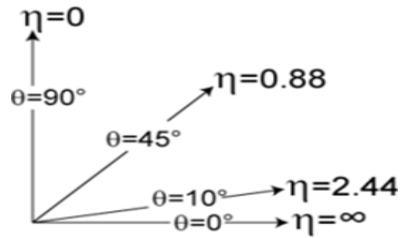

**FIG. 7**: Definition of Pseudorapidity

The previously published OPAL lower limit on the charged Higgs-boson mass, under the assumption of BR($H^\pm \to \tau\nu_\tau$) + BR($H^\pm \to q\bar{q}$) =1 is $m_{H^\pm}$ > 59.5GeV at $\sqrt{s}$ ≤ 183 GeV[20,21]. Lower bounds of 74.4 – 79.3 GeV have been reported by the other LEP collaborations [22, 23]. The DELPHI Collaboration also constrained the charged Higgs-boson mass in 2HDM [24] to be is $m_{H^\pm}$ > 76.7GeV GeV. Also a searches for the charged Higgs boson have been performed at $\sqrt{s}$ = 1.8 TeV in the $\tau + jets + \not{E}_T + \ell$ and $\tau$ lepton decays to hadrons where $\ell$ = e or $\mu$ in [25] and $\ell$ = e, $\mu$ or $\tau$ in [26].

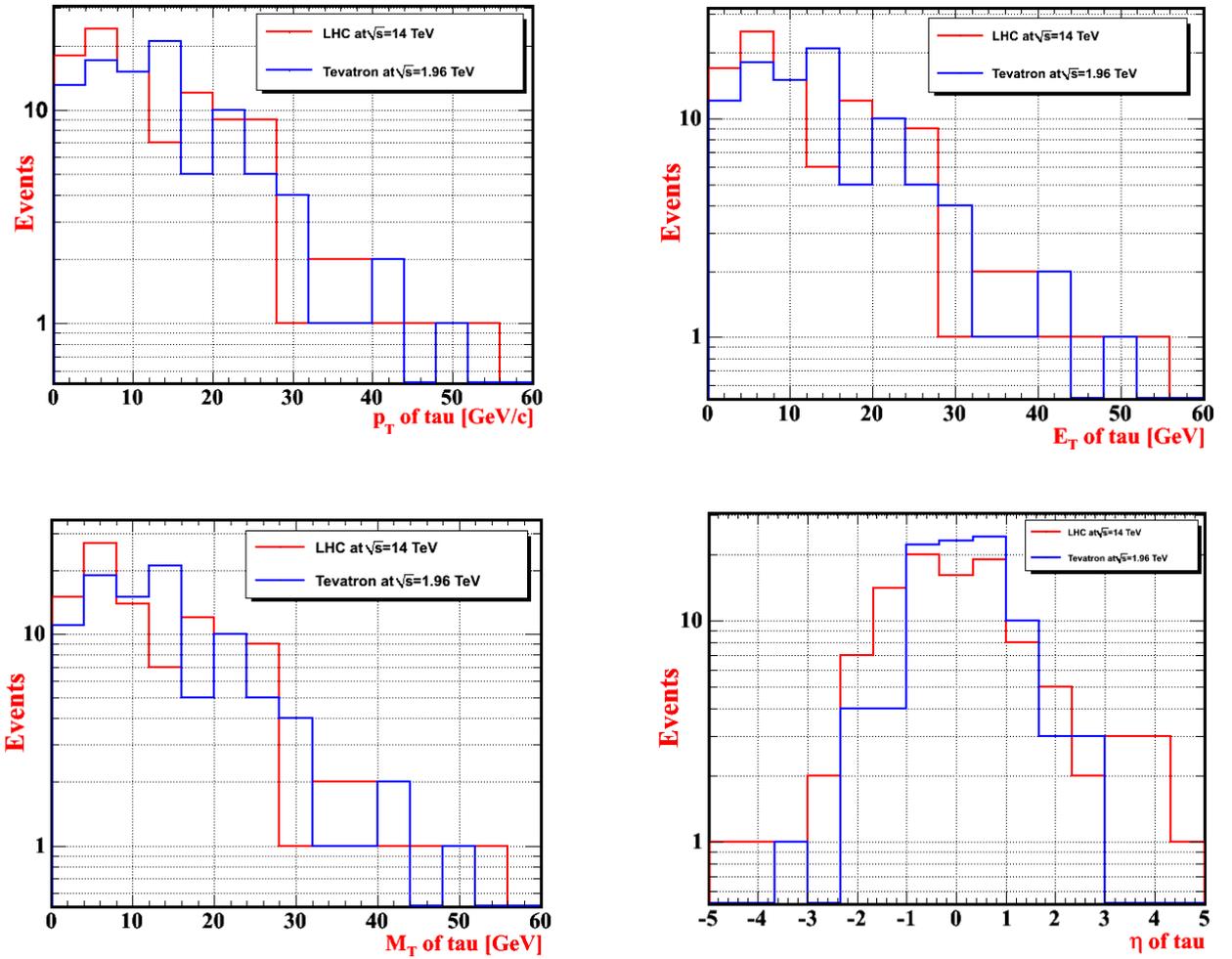

**FIG. 8**: Transverse momentum, Transverse energy, Transverse mass and pseudorapidity of tau produced from decay of charged Higgs at the LHC 14 TeV and at Tevatron 1.96 TeV using Pythia8.

## 2.5 Neural Networks Discrimination

The neural network method used for b-tagging in the OPAL SM Higgs-boson search [27] is used to calculate the discriminating the charged Higgs signal from the SM background. The inputs to the neural network include information about the electrons and the muons as transverse momentum, transverse mass and pseudorapidity. The main background in this search comes from decay of $w^{\pm}$ to electron and muons. The signal depends on the Higgs-boson masses and is very clean via electron and muon in the event. For purely leptonic events the first two candidates were retained and the rest were neglected as $\tau$ particles. For semileptonic events, only the first one was retained as a $\tau$ candidate. The resulting samples are completely dominated by background, the contribution of a Higgs signal being at most 0.5%.The statistical analysis is based on weighted event counting, with the weights computed from physical observables, also called discriminating variables of the candidate events. An improved analysis has been designed for the fully leptonic channel where BR ($H^+ \to \tau^+\tau_v$) = 1 and the rejection of the $w^+w^-$ background has been refined with Artificial Neural Networks (ANNs) discrimination. When dealing with semileptonic final states, the $\tau$ candidate jet definition was refined removing particles that were not likely to come from $\tau$ decay.

In current work we designed an artificial neural network consists of 4 layers figure 9. The first layer is the *input layer* and consists of 3 neurons (the neuron is the processing unit), the 3 neuron receive the input variables of a particle to the neural network figure 10 (Transverse momentum npt, transverse mass *nmt* and pseudorapidity *neta*). The second layer is a *hidden layer* consists of 5 neurons and the third layer also is a *hidden layer* consists of 3 neurons. The fourth layer is the *output layer* and consists of one neuron which gives the type of the particle gives 1 for the signal and 0 to the background as shown in figure 9. We trained and tested the neural network using two samples from the signal and the background and the signal sample consists of two sets, one from the LHC and the other from the Tevatron. The events of the signal and the background which we used it are stored in the Tree of ROOT data analysis.

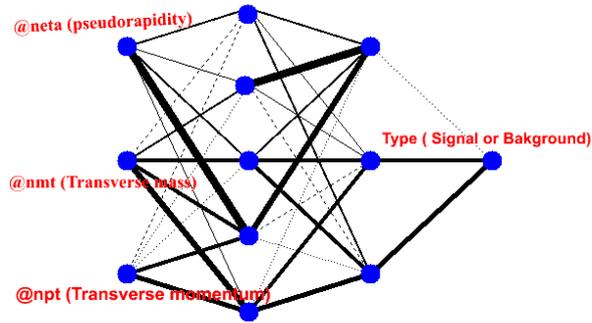

**FIG. 9**: Structure of the neural network where input layer consists of 3 neurons (transverse momentum, transverse mass and pseudorapidity) , output layer with one neuron and two hidden layers with 8 neurons.

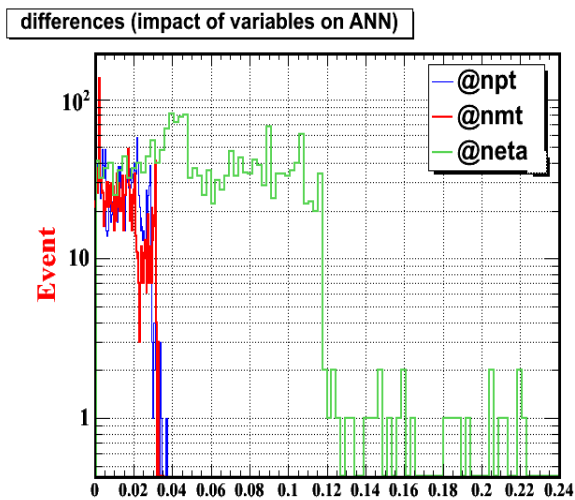

**FIG. 10**: How each variable influences the network (npt is transverse momentum, nmt is the transverse mass and neta is the pseudorapidity)

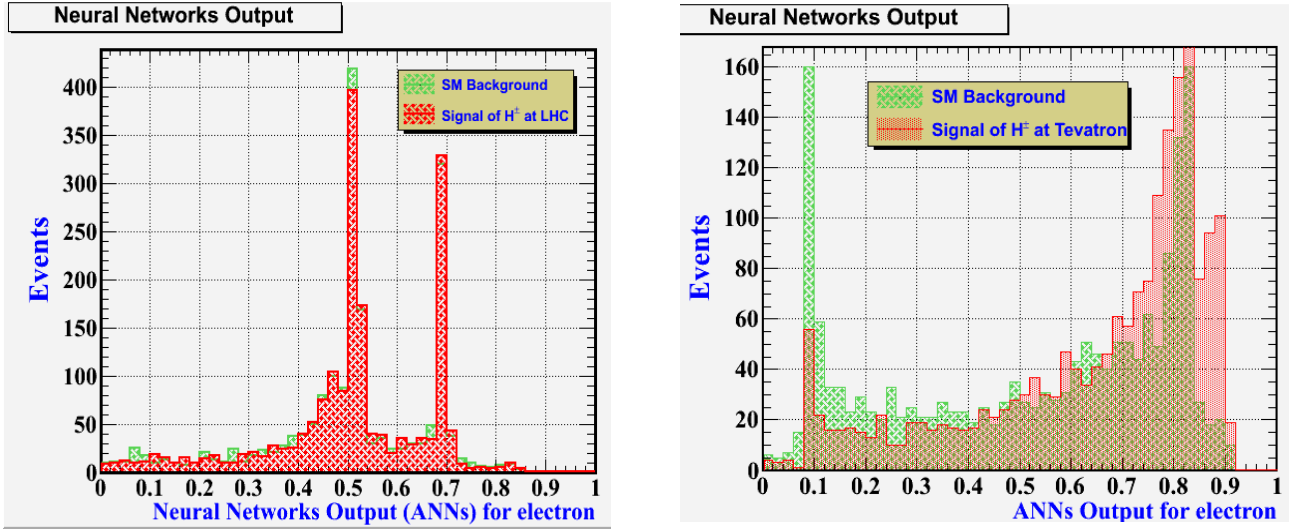

**FIG. 11**: Final discrimination of Neural Network outputs for decay $H^{\pm}$ of 120 GeV to electron in final state versus background plot (left) at the LHC 14 TeV Plot (right) at Tevatron 1.96 TeV.

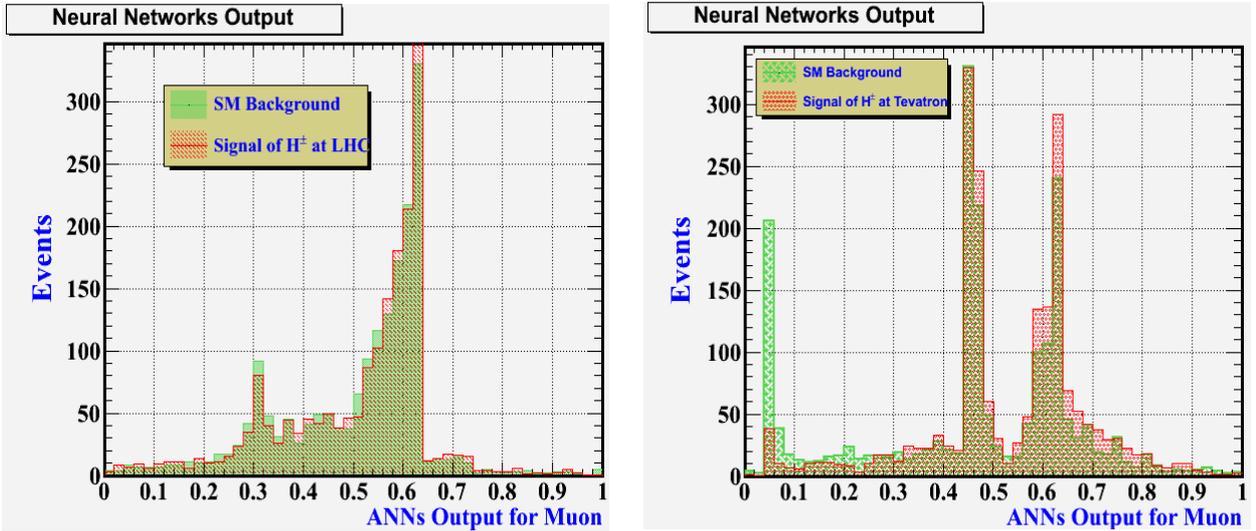

**FIG. 12**: Final discrimination of Neural Network outputs for decay $H^{\pm}$ of 120 GeV to muon in final state versus background Plot (left) at the LHC 14 TeV Plot (right) at Tevatron 1.96 TeV.

one lepton, having $E_T > 10$ GeV (electron) or $p_T > 15$ GeV (muon) one $\tau$ jet having $p_T > 25$ GeV and an electric charge opposite to that of the lepton. at least two jets having $p_T > 15$ GeV including at least one b-tagged jet. $E_T^{miss}$ is used as the discriminating variable to distinguish between SM $t\bar{t}$ events and those where top quark decays are edited by a charged Higgs boson, in which case the neutrinos are likely to carry away more energy $m_T = \sqrt{2p_T^\tau E_T^{miss}(1 - \cos\Delta\phi_{\tau,miss})}$

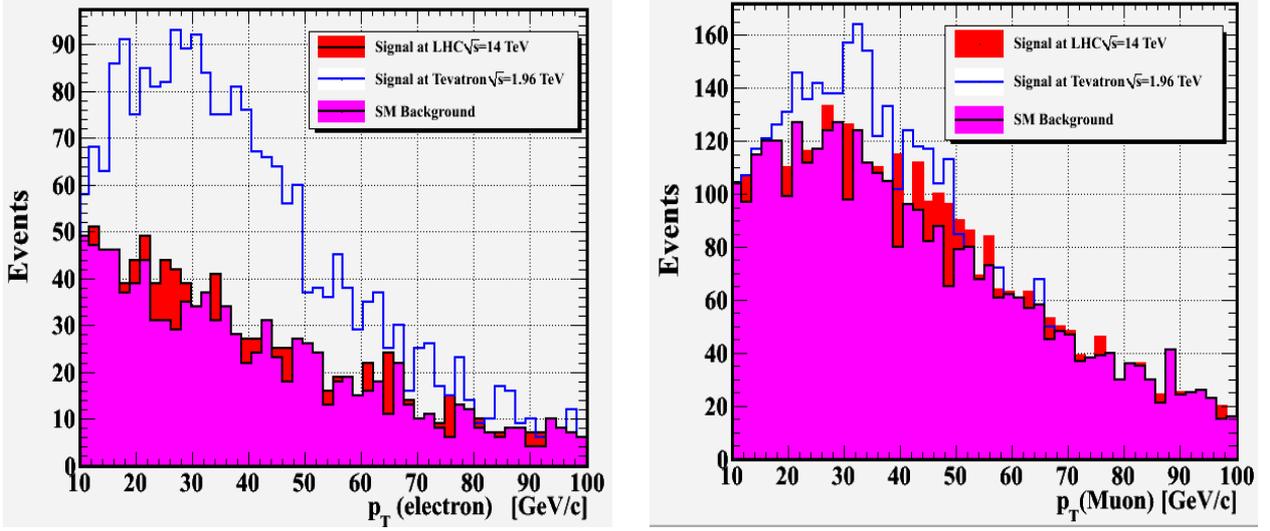

**FIG. 13**: Comparison of transverse momentum of electron and muon produced from decay of charged Higgs at the LHC 14 TeV and at Tevatron 1.96 TeV versus the SM background.

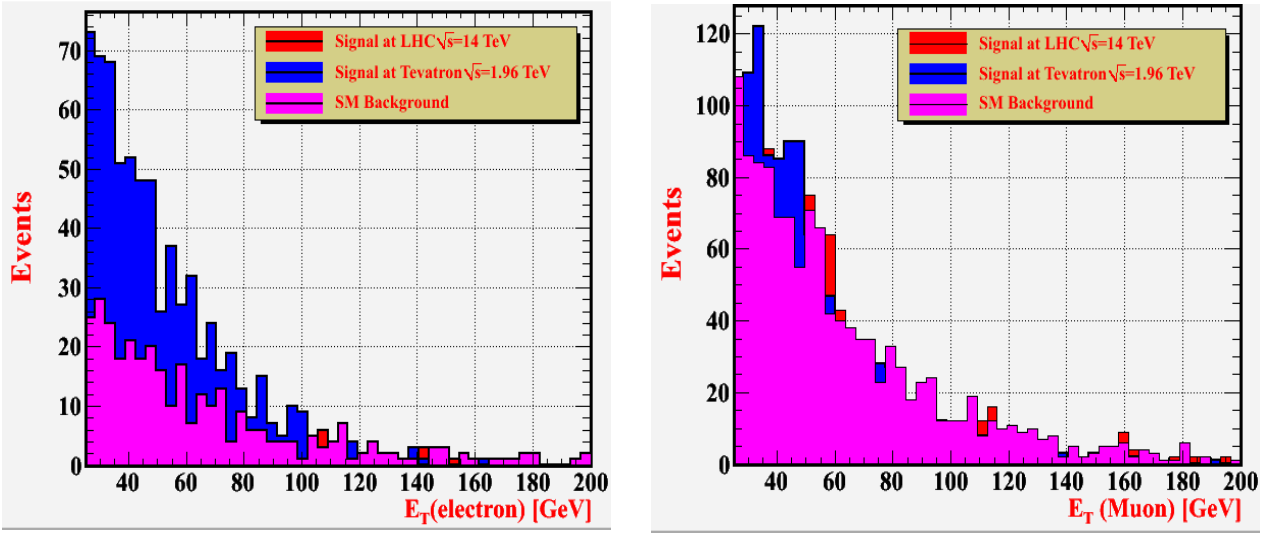

**FIG. 14**: Comparison of transverse energy of electron and muon produced from decay of charged Higgs at the LHC 14 TeV and at Tevatron 1.96 TeV versus the SM background.

The composition of the background depends on the targeted Higgs-boson mass region. In the low-mass selection the Higgs bosons are boosted and therefore the final state is electron and muon with the largest background contribution coming from decay of W boson as shown in figure 1. In some parts of the 2HDM parameter space both the fermionic $H^{\pm} \to \tau\nu$ and the bosonic decay modes contribute. Only the hadronic decays of is considered. Thus the events contain a tau lepton, electron, two jets and missing energy. Separating the signal from the $w^{\pm}$ background becomes difficult close to $m_{H^\pm} = m_{W^\pm}$ the preselection is designed to identify hadronic events containing a tau lepton plus significant missing energy and transverse momentum from the undetected neutrino.

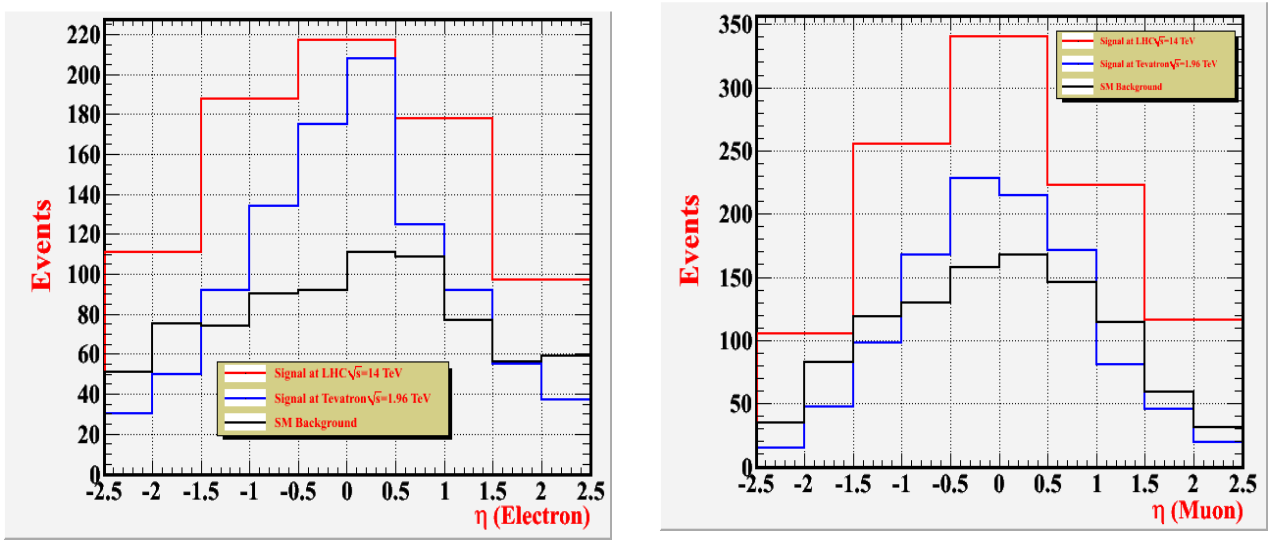

**FIG. 15**: Comparison of pseudorapidity of electron and muon produced from decay of charged Higgs at the LHC 14 TeV and at Tevatron 1.96 TeV versus the SM background.

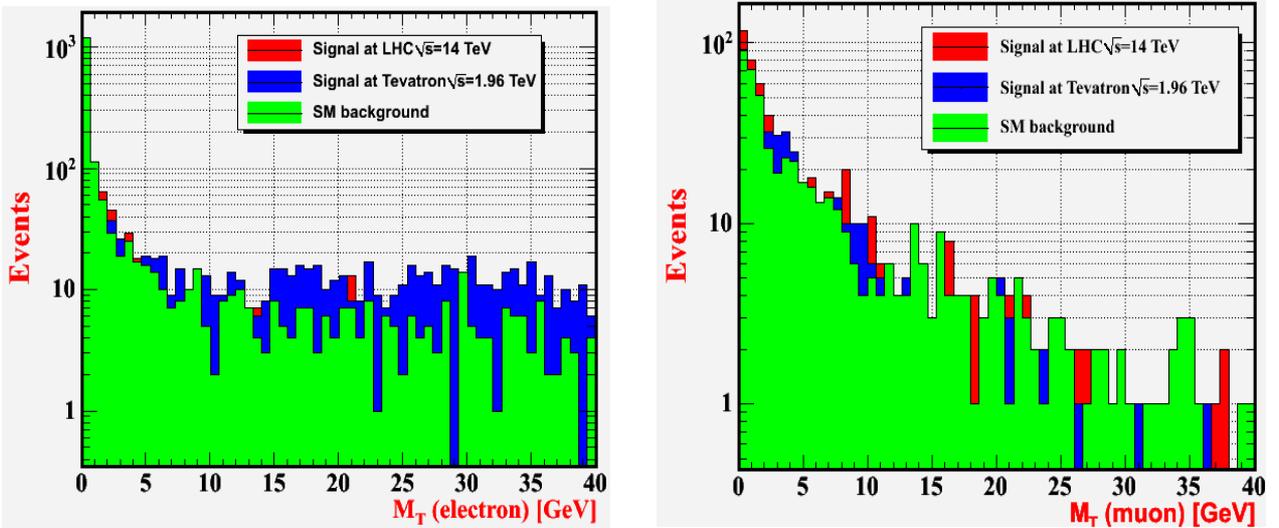

**FIG. 16**: Comparison of transverse mass of electron and muon produced from decay of charged Higgs at the LHC 14 TeV and at Tevatron 1.96 TeV versus the SM background.

The missing energy and momentum from *e.g.* tau charged Higgs boson decays are determined with [28] obtained by fulfilling the constraint $(p^{miss} + p^l + p^b)^2 = m_{top}^2$ on the leptonic side of lepton+jets $t\bar{t}$ events. More than one neutrino accounts for the invisible four-momentum $p^{miss}$ and its transverse component $\vec{P}_T^{miss}$. By construction, $m_T^H$ gives an event-by-event lower bound on the mass of the leptonically decaying charged (W or Higgs) boson produced in the top quark decay and it can be written as [29]:

$$(m_T^H)^2 = (\sqrt{m_{top}^2 + (\vec{p}_T^{\,l} + \vec{p}_T^{\,b} + \vec{p}_T^{\,miss})^2} - p_T^b)^2 - (\vec{p}_T^{\,l} + \vec{p}_T^{\,miss})^2$$

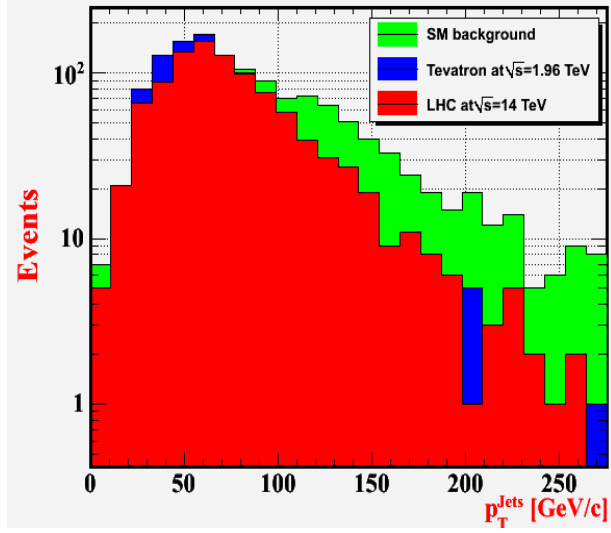

**FIG. 17**: Transverse momentum of jets produced in the final state at the LHC 14 TeV and at Tevatron 1.96 TeV versus the SM background.

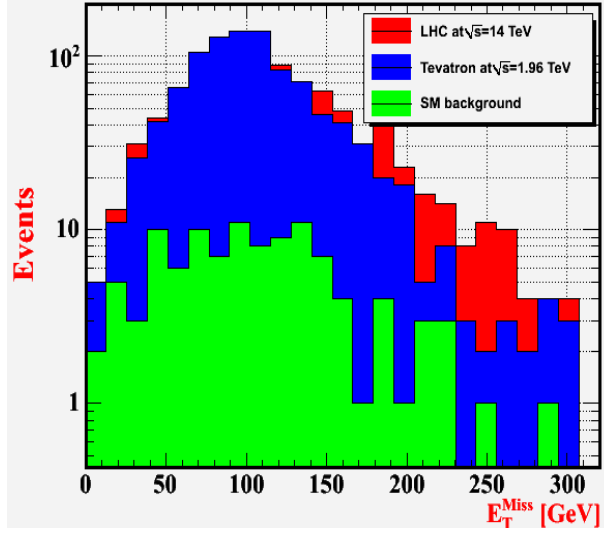

**FIG. 16**: The missing energy at the LHC 14 TeV and at Tevatron 1.96 TeV versus the SM background.

## CONCLUSION

We presented the results of a search for charged Higgs bosons ranging from 80 to 160 GeV. This analysis is based on Monte Caro simulation data and new discrimination technique is Artificial Neural Networks (ANNs) in the context of Two Doublet Higgs Model (2HDM) at both the LHC-CERN (ATLAS and CMS detectors) with proton-proton collisions at √s = 14 TeV and the Tevatron-Fermi Lab. (CDF and D0 detectors) with proton-antiproton collisions at √s = 1.96 TeV using top quark pair events with a electron-muon + Jets + missing energy ($\mu^+\nu_\mu\bar{\nu}_\tau\nu_\tau b e^-\bar{\nu}_e\nu_\tau\bar{\nu}_\tau\bar{b}$) in the final state and we assumed that the branching ratio of the charged Higgs boson to a $\tau$ lepton and a neutrino assumed Br ($H^+ \rightarrow \tau^+\tau_\nu$) = 100%.

# AKNOWLEDMENT

It is a pleasure to thank Prof. Torbjorn Sjostrand, Department of Theoretical Physics, Lund University, Lund, Sweden and the main author of Monte Carlo Event Generator (MCEG) Pythia8 for useful discussions.